# Single-Layer ScI$_2$: A Paradigm for Valley-Related Multiple Hall Effect

Zhonglin He, Rui Peng, Ying Dai*, Baibiao Huang and Yandong Ma*

School of Physics, State Key Laboratory of Crystal Materials, Shandong University, Shandanan Street 27, Jinan 250100, China

*Corresponding author: daiy60@sina.com (Y.D.); yandong.ma@sdu.edu.cn (Y.M.)

Valley-related multiple Hall effect in two-dimensional lattice is of notable interest both for its fundamental physics and for its potential applications. In this work, by means of a low energy k·p model analysis, a mechanism of producing valley-related multiple Hall effect in hexagonal lattice via strain engineering is proposed, and a general picture of valley-contrasted band inversion is developed. Through first-principles calculations, this mechanism is further established in a ferromagnetic hexagonal lattice of single-layer ScI$_2$. Single-layer ScI$_2$ prefers in-plane magnetization and exhibits neither anomalous valley Hall effect nor valley-polarized quantum anomalous Hall effect in nature. Remarkably, these two Hall effects emerge simultaneously in this system under 4.705% tensile strain and disappear simultaneously when further increasing strain, suggesting the exotic valley-related multiple Hall effect. The underlying physical mechanism is revealed using a model analysis and is generally applicable. Our work greatly enriches the valley-related physics.

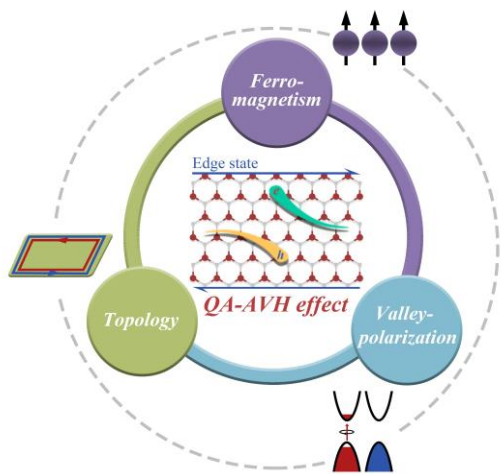

## Introduction

Spin-orbit coupling (SOC) characterizes the relativistic interaction between spin and momentum degrees of freedom of electrons, while time-reversal symmetry (*T*) describes the symmetry of physical law under time reversal transformation. The interplay between SOC and *T*-breaking is central to the rich phenomena observed in condensed matter systems.[1-5] In recent years, new phases of matter emerge from coupling this interplay with two-dimensional (2D) hexagonal lattices.[6-9] Anomalous valley Hall (AVH) effect is one of the defining features of such coupling, which is dictated by the absence of inversion symmetry (*P*).[7] The appearance of AVH effect is of intrinsic interest as it allows for the direct detection of valley current and therefore could be useful for valleytronics.[10,11] Valley-polarized quantum anomalous Hall (VP-QAH) effect, which is signaled by quantized charge Hall conductance and chiral edge states, is another defining feature.[12] Despite the insulating bulk states, the chiral edge states could conduct the charge current without dissipation, highly promising for spintronic and quantum computational devices.[1,13]

Compared with AVH and VP-QAH effects, their simultaneously existence in a ferromagnetic hexagonal lattice is more intriguing and of great importance because of its novel physics and multifarious potential applications.[14] Yet, the valley-related multiple Hall effect is rare in nature and its design is proved unexpectedly difficult[15], although exciting progress has been made in discovering both AVH and VP-QAH effects separately.[16-21] Physically, this challenge relates to the inherent exclusion of AVH and VP-QAH effects in 2D hexagonal lattices. While the former generally forms in 2D semiconductors with a wide band gap at the valley, the latter favors to occur in narrow-gap 2D materials. The appearance of VP-QAH effect associated to valley-dependent band inversion typically accompanies with the deformation of AVH effect and vice versa.[14,15] Actually, up to now, the simultaneous appearance and disappearance of valley-related multiple Hall effect in a ferromagnetic hexagonal lattice has not been reported[14,15,18-21], and the physics remains unknown.

Here, using a low energy k·p model, we propose a mechanism of capturing valley-related multiple Hall effect in ferromagnetic hexagonal lattice by introducing strain, and map out a general picture of valley-contrasted band inversion. With first-principles evidences, we further demonstrate this mechanism in a ferromagnetic hexagonal lattice of single-layer $ScI_2$. Single-layer $ScI_2$ favors in-plane magnetization and possesses neither AVH effect nor VP-QAH effect in nature. Intriguingly, these two Hall effects occur simultaneously in single-layer $ScI_2$ under 4.705% tensile strain and vanish completely when further increasing strain, establishing the simultaneous appearance and disappearance of valley-related multiple Hall effect. The underlying physics is unveiled through a model analysis, which is generally applicable. These insights will enrich the valley-related physics and guide the search for valley-related multiple Hall effect.

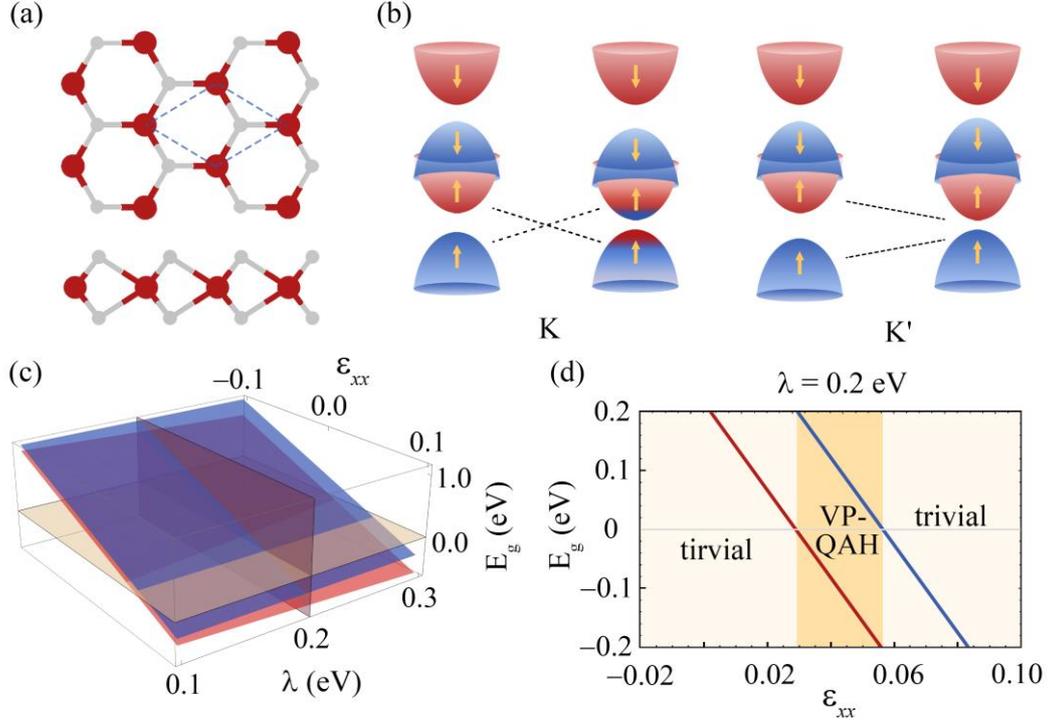

**Fig. 1** (a) Ferromagnetic hexagonal lattice of $MX_2$ from top and side views. Red and grey balls in a indicate M and X atoms, respectively, and the unit cell is marked by dashed lines. (b) Schematic illustrations of bands around the K and K' valleys without (left panel) and with (right panel) introducing biaxial strain. In b, band and spin indices are characterized by red/blue colors and up/down arrows, respectively. (c) Evaluation of band gaps at the K and K' points as a function of SOC strength and biaxial strain obtained from the k·p model. Red and blue planes in c represent the band gaps at the K and K' points, respectively, and the yellow horizontal plane indicates the energy of 0 eV. (d) Band gaps at the K and K' points as a function of biaxial strain when setting SOC strength to 0.2 eV. Blue and red lines in d represent the band gaps at the K and K' points, respectively. The values of the parameters are shown in Table S1.

## Results and discussion

**Four-band k·p model for valley-related multiple Hall effect in ferromagnetic hexagonal lattice**

Considering valley-polarized physics is mainly concentrated on ferromagnetic hexagonal lattice with a formula of $MX_2$ (M: transition metal; X: non-metal elements),[7,14-17] we start from this lattice, as shown in Fig. 1a. Without losing of generality, we assume the energy extrema of conduction and valence bands locating at the K and K' points. Given $C_{3h}$ point group of wave vectors at the K and K' points, the symmetry adapted basis functions are

$$|\psi_u^\tau\rangle = |d_{z^2}\rangle,$$

$$|\psi_l^\tau\rangle = \frac{1}{\sqrt{2}}\left(|d_{xy}\rangle + i\tau|d_{x^2-y^2}\rangle\right).$$

Here, $\tau = \pm 1$ represents K/K' valley. The subscripts $u$ and $l$ are adopted to describe the upper and lower bands, respectively. The two-band k·p Hamiltonian around K and K' valleys has the form

$$H_0 = \frac{\Delta}{2}\hat{\sigma}_z + at(\tau\hat{\sigma}_x k_x + \hat{\sigma}_y k_y).$$

Here, $\hat{\sigma}_i$ ($i = x, y, z$) denotes the Pauli matrices for the two basis functions, $\Delta$ is the energy gap, $a$ is the lattice constant and $t$ represents the effective nearest neighbor hopping. Then the effective Hamiltonian of the four-band k·p model for the ferromagnetic hexagonal lattice with considering SOC and exchange interaction can be written as

$$H_{eff} = \hat{s}_0 H_0 + \lambda\tau\hat{s}_z \frac{\hat{\sigma}_z + \hat{\sigma}_0}{2} - \hat{s}_z \frac{m_1\hat{\sigma}_z + m_2\hat{\sigma}_0}{2}.$$

Here, $\hat{s}_i$ ($i = x, y, z$) denotes the Pauli matrices for spin. $\lambda$, $m_u\,(=\frac{m_1-m_2}{2})$ and $m_l\,(=\frac{m_1+m_2}{2})$ describe spin splitting caused by SOC and exchange interaction, respectively. Upon considering spin, the basis functions transform into $|\psi_l^\tau\uparrow\rangle, |\psi_u^\tau\uparrow\rangle, |\psi_l^\tau\downarrow\rangle$ and $|\psi_l^\tau\uparrow\rangle$. By diagonalizing the effective Hamiltonian, the energy spectra can be given as

$$E_i^s(k_x, k_y) = \frac{\Delta + i(m_1 - \lambda\tau) + s\epsilon_i(k_x, k_y)}{2}.$$

Here, $i = \pm 1$ (+1 for upper band $u$ and -1 for lower band $l$) and $s = \pm 1$ represent the band and spin indices, respectively. As schematically illustrated in Fig. 1b, this gives rise to a spontaneous valley polarization of $2\lambda$, which can generate the intriguing AVH effect.

The band gap at K/K' valleys is $E_g^\tau = E_{i=1}^{s=1}(\tau) - E_{i=-1}^{s=1}(\tau) = \Delta + m_2 - \lambda\tau$. For the hexagonal lattice that does not have band inversion at both valleys, the following expression can be obtained

$$(\Delta + m_2 + \lambda)(\Delta + m_2 - \lambda) > 0.$$

Obviously, the increase of SOC strength would decrease the band gap at K valley and increase the gap at K' valley, indicating the possibility of introducing valley-dependent band inversion as well as VP-QAH effect. However, SOC strength within the M-$d$ orbitals is generally not strong enough to induce band inversion because the ferromagnetic hexagonal lattice of MX$_2$ normally exhibits a moderate band gap. Therefore, additional strategies should be explored. Here, strain engineering is employed to induce valley-dependent band inversion. According to deformation potential theorem,[22,23] the deformation term induced by small biaxial strain can be expressed as

$$H_\varepsilon = \frac{D(\varepsilon_{xx}+\varepsilon_{yy})\hat{s}_0\hat{\sigma}_z}{4} = \frac{D\varepsilon_{xx}\hat{s}_0\hat{\sigma}_z}{2},$$

where $D$ is deformation potential. $\varepsilon_{xx}/\varepsilon_{yy}$ represents strain along $x/y$ axis and is defined as $\varepsilon_{xx} = \frac{a-a_0}{a}$, where $a_0/a$ is the lattice parameter without/with strain engineering. In this case, the band gap at K/K' valley is expressed as

$$E_g^\tau = \Delta + m_2 - D\varepsilon_{xx} - \lambda\tau.$$

Based on this relationship, we plot the band gap $E_g^\tau$ as a function of SOC strength and biaxial strain in Fig. 1c. For realizing valley-dependent band inversion, the band gaps at K and K' valleys should have the relation $E_g^{\tau=1} \cdot E_g^{\tau=-1} < 0$. From Fig. 1c, it can be seen that, with increasing $\varepsilon_{xx}$, energy surfaces of $E_g^{\tau=1}$ and $E_g^{\tau=-1}$ both decrease, and intersect with the energy surface of $E_g = 0$ in sequence. This indicates the valley-dependent band inversion, as indicated in Fig. 1b, and thus the VP-QAH effect can be realized via strain engineering.

To illustrate this phenomenon more clearly, we plot the topological phase diagram at $\lambda = 0.2$ eV as a function of strain in Fig. 1d. As shown in Fig. 1d, under strain less than 0.029, the band orders are normal at both valleys, indicating a trivial phase. Upon increasing strain to 0.029, band inversion occurs at K valley, while the band order at K' valley remains normal, forming a VP-QAH phase. When further increasing strain to 0.057, band inversion occurs at K' valley, which indicates a trivial state. To confirm these phases, we calculate the Chern numbers $c(\varepsilon_{xx})$ at $\varepsilon_{xx}$ = 0.000, 0.040 and 0.070 which are estimated to be $c(0.000)$ = $c(0.070)$ and $c(0.040) = 1$. Therefore, with increasing strain, the hexagonal lattice transforms from trivial to VP-QAH phases, and then back to trivial state. We also calculate the valley Chern number $c_v$ of the resultant VP-QAH state and find $c_v$ is nearly zero. This suggest that the realization of VP-QAH effect does not deform the valley physics as well as the AVH effect, thus holding the possibility for achieving valley-related multiple Hall effect.

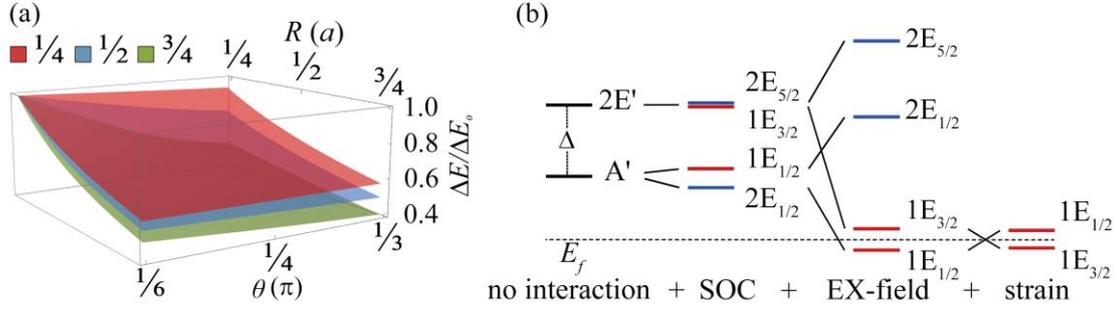

**Fig. 2.** (a) Variation of $\Delta E/\Delta E_0$ as a function of $\theta$ and $R$ with setting $\rho/R = 1/4$ (red), 1/2 (blue) and 3/4 (green). (b) Diagram of orbital evaluations for the two low energy bands at the K point with including SOC, exchange field and biaxial strain in sequence. Red and blue lines in b correspond to the spin up and down states, respectively. The Fermi level is indicated by the dashed line. The IRs of states are labelled using Mülliken notations.

To get further insight into the physical mechanism of the VP-QAH effect induced by strain, we employ the crystal field theory. The Hamiltonian for electrons of M cation consists of two terms $H = H_0 + V$, where $H_0$ is the Hamiltonian for the free cation and $V$ is the potential provided by its ligands[24]. Considering $D_{3h}$ symmetry of the hexagonal lattice, for $d$-levels, $V$ can be expanded as

$$V(\vec{r}) = A_{00}Y_{00} + A_{20}r^2Y_{20} + A_{40}r^4Y_{40}.$$

Here, $Y_{lm} = Y_{lm}(\theta,\phi)$ is a normalized spherical harmonic, and $A_{lm} = -\dfrac{4\pi}{2l+1}\sum_j \dfrac{q_j Y_{lm}^*(\theta_j,\phi_j)}{R_j^{l+1}}$. $q_j$ and $(R_j,\theta_j,\phi_j)$ represent the charge and location of the $j$-th ligand [see Fig. S1, Supporting Information]. The crystal field splitting between the symmetric basis atomic functions $\phi_\gamma = d_{xy} + i\gamma d_{x^2-y^2}$ ($\gamma = \pm 1$) and $\phi_0 = d_{z^2}$ can be expressed as

$$\Delta E = \langle \phi_\gamma | V(\vec{r}) | \phi_\gamma \rangle - \langle \phi_0 | V(\vec{r}) | \phi_0 \rangle.$$

Given the localized character of $d$ orbitals, the matrix element can be simplified as

$$\langle \phi_\gamma | V(\vec{r}) | \phi_\gamma \rangle \approx \int d\Omega \int_0^\rho \phi_\gamma^*(\vec{r}) V(\vec{r}) \phi_\gamma(\vec{r}) dr,$$

where $\rho$ is the upper limit of integral. Owing to $D_{3h}$ symmetry, it's reasonable and convenient to set $\theta_j = \theta$ ($j$=1,2,3), $(\pi - \theta_j) = \theta$ ($j$=4,5,6) and $R_j = R$. By increasing tensile strain, both $\theta$ and $R$ increase. Fig. 2a illustrates the variation of $\Delta E/\Delta E_0$ as a function of $\theta$ and $R$ with setting $\rho/R$=1/4, 1/2 and 3/4, where $\Delta E_0 = \Delta E(\theta = \pi/4, R = a/4)$. Obviously, $\Delta E$ shrinks with increasing strain. As shown in Fig. 2b, when excluding SOC and exchange interaction, the irreducible representations (IRs) at the K point are 2E' and A' for $|\psi_u^{\tau=1}\rangle$ and $|\psi_l^{\tau=1}\rangle$, respectively. The inclusion of tensile strain weakens the crystal field splitting between $\phi_\gamma$ and $\phi_0$, corresponding to the narrowing of the gap between the Bloch states $|\psi_u^{\tau=1}\rangle$ and $|\psi_l^{\tau=1}\rangle$. With including SOC, the two one-dimensional IRs split into four IRs of 1E$_{3/2}$, 2E$_{5/2}$, 2E$_{1/2}$ and 1E$_{1/2}$, which respectively relate to the four basis functions $|\psi_u^{\tau=1}\uparrow\rangle, |\psi_u^{\tau=1}\downarrow\rangle, |\psi_l^{\tau=1}\downarrow\rangle$ and $|\psi_l^{\tau=1}\uparrow\rangle$. Upon further including exchange field (EX-field), the two states with IRs 1E$_{3/2}$ and 1E$_{1/2}$ locate on the two sides of the Fermi level. Apparently, 1E$_{3/2}$ and 1E$_{1/2}$ state would get close and even inversed when applying tensile strain. Thus, band inversion can be introduced through tensile strain.

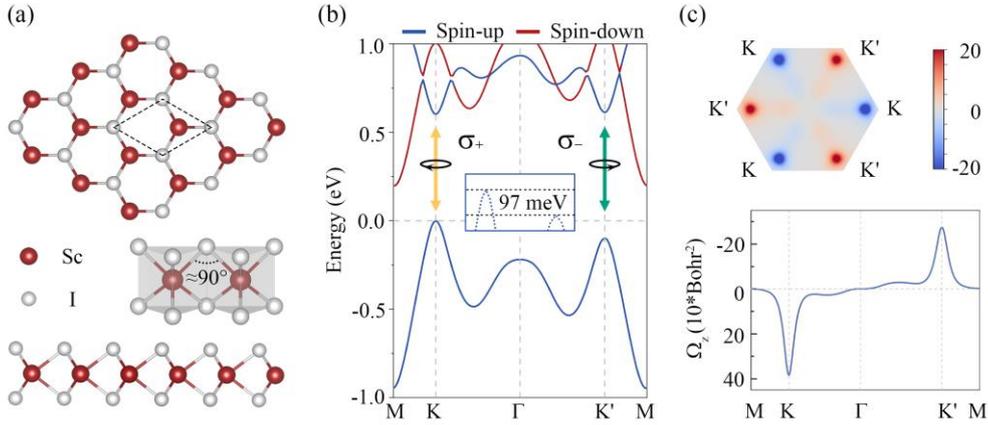

**Fig. 3**. (a) Crystal structure of single-layer ScI$_2$ from top and side views. (b) Spin-polarized band structure of single-layer ScI$_2$ with considering SOC and out-of-plane magnetization. (c) Berry curvature of single-layer ScI$_2$ as a counter map over the 2D Brillouin zone and as a curve along the high-symmetry points. The Fermi level is set to 0 eV.

**Valley-related multiple Hall effect in single-layer ScI$_2$**

After establishing the mechanism of capturing valley-related multiple Hall effect in the ferromagnetic hexagonal lattice, we next discuss its realization in a real material of single-layer ScI$_2$. Fig. 3a presents the crystal structure of single-layer ScI$_2$. It exhibits a hexagonal lattice with the space group of $P\bar{6}m2$, satisfying the prerequisite of *P*-breaking. The lattice constant of single-layer ScI$_2$ is optimized to be 3.98 Å, agreeing well with previous work.[25] The stability of single-layer ScI$_2$ is confirmed by phonon calculations and AIMD simulations, see Fig. S2a, b, Supporting Information.

Sc atom harbors a valence electronic configuration of 3$d^1$4$s^2$. After donating two electrons to the surrounded I atoms, only one valence electron left, resulting in a magnetic moment of 1 $\mu_B$ per unit cell for single-layer $ScI_2$. To estimate the magnetic ground state of single-layer $ScI_2$, three magnetic configurations are considered, including ferromagnetic (FM), antiferromagnetic (AFM) and Néel antiferromagnetic (NAFM) configurations. The FM configuration is found to be 199 and 185 meV per unit cell lower in energy than the AFM and NAFM configurations, respectively, satisfying condition of *T*-breaking. The FM ground state of single-layer $ScI_2$ is related to its crystal structure. As is shown in Fig. 3a, the Sc-I-Sc bonding angle is 83.79°, close to 90.0°. According to the Goodenough-Kanamori-Anderson rules[26-28], FM coupling should dominate the exchange interaction. To investigate the magnetization easy axis of single-layer $ScI_2$, we investigate its magnetic anisotropy energy (MAE). The MAE is calculated to be 0.42 meV per unit cell, indicating that single-layer $ScI_2$ prefers in-plane magnetization in nature.

Fig. S3a, Supporting Information, presents the band structure of single-layer $ScI_2$ without considering exchange interaction and SOC. Due to the half-occupied *d* orbital, there is a spin-degenerate band crossing the Fermi level. When considering exchange interaction but without SOC, as shown in Fig. S3b, Supporting Information, the spin degeneracy is lifted, resulting in a semiconducting feature with an indirect band gap of 0.25 eV. Interestingly, its band edges at the K and K' points generate a pair of valleys in both the conduction and valence bands. By further introducing SOC, as shown in Fig. S3c, Supporting Information, the valley degeneracy is preserved, indicating the valley polarization is absent in single-layer $ScI_2$, which can be attributed to its in-plane magnetization orientation. This means valley-polarized physics, like AVH and VP-QAH effects, are absent in single-layer $ScI_2$ in nature. To explore the valley-polarized physics, we suppose that the magnetization orientation is tuned from in-plane to out-of-plane in the following. As we will show later, strain engineering can modulate the magnetization orientation of single-layer $ScI_2$. The band structure of single-layer $ScI_2$ with SOC and out-of-plane magnetization is shown in Fig. 3b. We can see K valley shifts above K' valley in the valence band, giving rise to the intriguing spontaneous valley polarization of 97 meV. While for the conduction band, the valley polarization is negligible, which results from the out-of-plane orbital contributions. The spontaneous valley polarization can lead to intriguing AVH effect, and also makes it possible for the valley-dependent band inversion.

To characterize the valley-contrasting physics in single-layer $ScI_2$, the calculation of its Berry curvature is performed according to the following expression[29]:

$$\Omega(k) = -\sum_n \sum_{n \neq n'} f_n \frac{2Im \langle \psi_{nk}|\hat{v}_x|\psi_{n'k}\rangle \langle \psi_{n'k}|\hat{v}_y|\psi_{nk}\rangle}{(E_n(k)-E_{n'}(k))^2}$$

Here $f_n$ is the Fermi-Dirac distribution function, $E_n(k)$ is eigenvalue of the Bloch state $\psi_{nk}$, and $\hat{v}_{x/y}$ is the velocity operator. As shown in Fig. 3c, there are two peaks at K and K' valleys, which have opposite signs and different absolute values. The former character arises from its *P*-breaking, while the later corelates with *T*-breaking. Under an in-plane electric field, Bloch electrons from the valleys in single-layer ScI$_2$ will acquire an anomalous velocity $v \sim E \times \Omega(k)$.[10,30] In this regard, by shifting the Fermi level between K and K' valleys in the valence band, the spin-down holes from K valley will accumulate at one boundary of the sample in the presence of an in plane electric field, producing the intriguing AVH effect.

Following the mechanism obtained from the k·p model, we then engineer the valley-related physics of single-layer ScI$_2$ trough introducing strain. The band structures of single-layer ScI$_2$ without including SOC and exchange interaction under various strain are presented in Fig. S4, Supporting Information, from which we can see that the crystal field splitting between $\phi_\gamma$ and $\phi_0$ obviously shrinks with increasing tensile strain. Fig. S5, Supporting Information, displays the band structures of single-layer ScI$_2$ with SOC under various strain. As expected, by increasing tensile strain, the K and K' valleys in the conduction band move down, while the K and K' valleys in the valence band move up. When the tensile strain reaches to 4.705%, the band gap at K valley is closed, while the gap at K' valley is preserved, giving rise to the half-valley-metal state [see in Fig. S6a, Supporting Information]. By further increasing the tensile strain, the band gap at K valley is reopened, yielding a band inversion at K valley. While the band order at K' valley remains normal, the VP-QAH phase is realized in single-layer ScI$_2$. Simultaneously, the spontaneous valley polarization is preserved, suggesting the appearance of the exotic valley-related multiple Hall effect. With increasing the tensile strain to 5.100%, the band gap at K' valley is closed, leading to another half-valley-metal state [see in Fig. S6b, Supporting Information]. Upon increasing the tensile strain larger than 5.100%, band inversion occurs at K' valley. Since band inversion already occurs at K valley, VP-QAH state transforms back into the trivial state.

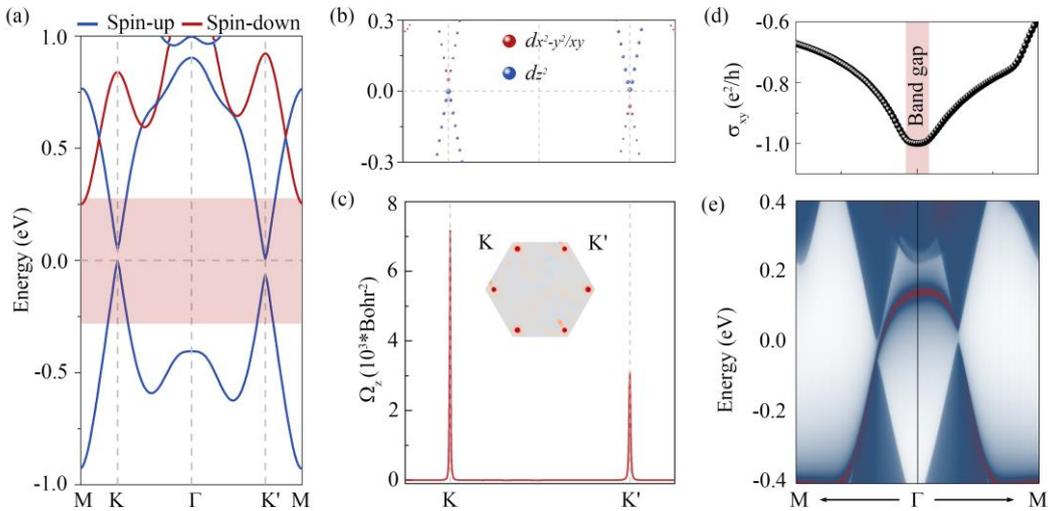

**Fig. 4** (a) Spin-polarized band structure of single-layer ScI$_2$ with considering SOC under 5% tensile strain. (b) Orbital-resolved bands near the Fermi level. (c) Berry curvature of single-layer ScI$_2$ as a curve along the high-symmetry points. Insert in c describes Berry curvature as a counter map over the 2D Brillouin zone, which indicates the nonzero Chern number for single-layer $ScI_2$. (d) Anomalous Hall conductivity versus chemical potential for single-layer $ScI_2$. (e) Edge spectra for the VP-QAH state of single-layer $ScI_2$. The Fermi level is set to 0 eV.

Taking the case under 5% tensile strain as an example, we investigate the valley-related multiple Hall effect in single-layer $ScI_2$. Fig. 4a illustrates the band structure of single-layer $ScI_2$ under 5% tensile strain. It can be seen that the spontaneous valley polarization in the valence band is preserved, which is estimated to be 63 meV. Such sizeable valley polarization is beneficial for the observation of AVH effect in single-layer $ScI_2$. In addition to the valence band, spontaneous valley polarization also occurs in the conduction band [Fig. 4a], which is found to be 49 meV. This suggests the appearance of orbitals with in-plane character in the conduction band, implying the valley-dependent band inversion. From the valley-selective circular dichroism and Berry curvature shown in Fig. S7b, Supporting Information and Fig. 4c, it can be seen that the band inversion indeed occurs at K valley.[31] And the band inversion at K valley is further confirmed by the orbital-resolved bands presented in Fig. 4b. These facts suggest that the VP-QAH effect is realized in single-layer $ScI_2$ under 5% tensile strain. To confirm this nontrivial phase, we compute the anomalous Hall conductance $\sigma_{xy}$ around the Fermi level [see Fig. 4d] and obtain the quantized anomalous Hall conductance $\sigma_{xy} = ce^2/h$ with a Chern number $c = 1$. Moreover, as is depicted in Fig. 4e, there is one chiral gapless edge mode connecting the valence and conduction bands within the bulk gap, which is a hallmark of VP-QAH phase. Therefore, the valley-related multiple Hall effect is realized in single-layer $ScI_2$.

**Strain engineering and physical mechanism**

It should be noted that the above results are based on the precondition of ferromagnetism with out-of-plane magnetization orientation. We then investigate the evaluation of magnetic properties of single-layer $ScI_2$ under strain engineering. Fig. S8a, Supporting Information, displays the variation of energy difference between different magnetic configurations as a function of strain, which shows that FM configuration becomes more stable with increasing strain. This can be ascribed to that the increasing of tensile strain forces Sc-I-Sc bonding angle closer to 90°. According to Goodenough-Kanamori-Anderson rules,[26-28] FM coupling would be enhanced. The evaluation of MAE under strain engineering is shown Fig. S8b, Supporting Information. Under strain below 4.705%, single-layer $ScI_2$ favors in-plane magnetization. By increasing strain larger than 4.705% (the critical point for band inversion at K point), out-of-plane magnetization is preferred. When further increasing the strain above 5.100% (the critical point for band inversion at K' point), in-plane magnetization becomes more favorable again. Note that in-plane magnetization would vanish valley-polarized physics, both AVH and VP-QAH effects are absent in single-layer $ScI_2$ under strain below 4.705%. For strain larger than 4.705%, AVH and VP-QAH effects appears simultaneous in single-layer ScI$_2$, giving rise to valley-related multiple Hall effect. And these two Hall effects disappear completely when strain is larger than 5.100%. Therefore, the simultaneous appearance and disappearance of valley-related multiple Hall effect in a ferromagnetic hexagonal lattice is realized.

At last, we address the underlying physical mechanism for the variation of MAE under strain. For MAE, it can be expressed as[32]

$$\mathrm{MAE} = -\xi_{Sc}^2 \sum_{o,u,\sigma_1,\sigma_2} \sigma_1 \sigma_2 \frac{\left|\langle u,\sigma_1|\hat{L}_z^{Sc}|o,\sigma_2\rangle\right|^2 - \left|\langle u,\sigma_1|\hat{L}_x^{Sc}|o,\sigma_2\rangle\right|^2}{E_{u,\sigma_1} - E_{o,\sigma_2}}$$

where $u$/$o$ refer to unoccupied/occupied state near the Fermi level, $E_{u/o,\sigma}$ is energy of the state, $\sigma_1/\sigma_2 = \pm 1$ is spin index, and $\xi_{Sc}$ represents the SOC strength of Sc atoms. The negative (positive) MAE indicates the preference of out-of-plane (in-plane) magnetization. For unstrained single-layer ScI$_2$, as shown in Fig. S3d, Supporting Information, $|u,\sigma_1\rangle = \alpha|\phi_0 \downarrow\rangle + \beta|\phi_\gamma \downarrow\rangle$ and $|o,\sigma_2\rangle = |\phi_\gamma \uparrow\rangle$, then MAE can be simplified as

$$\mathrm{MAE} \approx \xi_{Sc}^2 \frac{\left|\langle \alpha\phi_0 + \beta\phi_\gamma|\hat{L}_z^{Sc}|\phi_\gamma\rangle\right|^2 - \left|\langle \alpha\phi_0 + \beta\phi_\gamma|\hat{L}_x^{Sc}|\phi_\gamma\rangle\right|^2}{\Delta}.$$

Here, $\Delta$ is the band gap. Because $\langle \phi_{j_1}|\hat{L}_z|\phi_{j_2}\rangle = 2i\delta_{j_1 j_2}$ and $\langle \phi_{j_1}|\hat{L}_x|\phi_{j_2}\rangle = 0$ ($j_1, j_2 = 0, \pm 1$),

$$\mathrm{MAE} \approx \xi_{Sc}^2 \frac{\left|\langle \beta\phi_\gamma|\hat{L}_z^{Sc}|\phi_\gamma\rangle\right|^2}{\Delta} > 0.$$

This indicates the in-plane magnetization for unstrained single-layer ScI$_2$, preventing the possibility of valley-polarized physics. By increasing strain larger than 4.705%, band inversion occurs at K valley. Then we have $|u,\sigma_1\rangle = \alpha_1|\phi_0 \uparrow\rangle + \beta_1|\phi_\gamma \uparrow\rangle$ and $|o,\sigma_2\rangle = \alpha_2|\phi_0 \uparrow\rangle + \beta_2|\phi_\gamma \uparrow\rangle$. And the MAE can be simplified as

$$\mathrm{MAE} \approx -\xi_{Sc}^2 \frac{\left|\langle \beta_1\phi_\gamma|\hat{L}_z^{Sc}|\beta_2\phi_\gamma\rangle\right|^2}{\Delta} < 0,$$

which is in favor of out-of-plane magnetization and valley-polarized physics. When further increasing strain larger than 5.100%, band inversion occurs at K' valley, and we have $|u,\sigma_1\rangle = |\phi_\gamma \uparrow\rangle$ and $|o,\sigma_2\rangle = |\phi_0 \uparrow\rangle$. The corresponding MAE is

$$\mathrm{MAE} \approx -\xi_{Sc}^2 \frac{\left|\left\langle\phi_\gamma\left|\hat{L}_z^{Sc}\right|\phi_0\right\rangle\right|^2 - \left|\left\langle\phi_\gamma\left|\hat{L}_x^{Sc}\right|\phi_0\right\rangle\right|^2}{\Delta} = 0.$$

Therefore, the SOC effect originates from I atoms should be taken into consideration.[33] The MAE transforms into

$$\mathrm{MAE} \approx -\xi_I^2 \frac{\left|\left\langle\phi_\gamma\left|\hat{L}_z^{I}\right|\phi_0\right\rangle\right|^2 - \left|\left\langle\phi_\gamma\left|\hat{L}_x^{I}\right|\phi_0\right\rangle\right|^2}{\Delta}.$$

Here, $\xi_I$ represents the SOC strength of I atoms. Because of the discrete rotational symmetry $C_3$, $L_z$ is a good quantum number, leading to a diminutive value of $\left|\left\langle\phi_\gamma\left|\hat{L}_z^{I}\right|\phi_0\right\rangle\right|$. On the contrary, $\hat{L}_x^{I}$ tends to mix the two states substantially, which indicates $\left|\left\langle\phi_\gamma\left|\hat{L}_x^{I}\right|\phi_0\right\rangle\right| \gg \left|\left\langle\phi_\gamma\left|\hat{L}_z^{I}\right|\phi_0\right\rangle\right|$. And the MAE is

$$\mathrm{MAE} \approx \xi_I^2 \frac{\left|\left\langle\phi_\gamma\left|\hat{L}_x^{I}\right|\phi_0\right\rangle\right|^2}{\Delta} > 0,$$

suggesting the preferred in-plane magnetization and the absence of valley-polarized physics.

## Conclusion

In conclusion, using a k·p model analysis and first-principles calculations, we introduce a mechanism of capturing valley-related multiple Hall effect in ferromagnetic hexagonal lattice, and further demonstrate this mechanism in a real material of single-layer ScI$_2$. The underlying physics is established and is generally applicable. Our results greatly enrich the valley-related physics and will guide the search for valley-related multiple Hall effect.

## Methods

### The DFT method and parameters

First-principles calculations are performed based on density-functional theory (DFT) as implemented in Vienna ab initio Simulation Package (VASP).[34-36] The exchange-correlation interaction is described by generalized gradient approximation (GGA) in the form of Perdew-Burke-Ernzerhof (PBE) functional.[37] The cutoff energy is set to 500 eV, and the convergence criterion of energy is set to $10^{-6}$ eV. Structures are fully relaxed until the force on each atom less than $10^{-2}$ eV/Å. A Monkhorst-Pack grid of 11 × 11 × 1 is used to sample the Brillouin zone. For MAE calculations, a dense grid of 15 × 15 × 1 is used. To avoid interactions between adjacent layers, a vacuum space of 30 Å is adopted. GGA + U method is employed to describe the strong correlated correction of Sc-3$d$ electrons.[38] Following the previous works, the U value is chosen as 3 eV.[39] Berry curvature and anomalous Hall conductivity are calculated with employing the WANNIER90 code.[40]

**Chern number and valley Chern number calculations**

The Chern number $c$ and the valley Chern number $c_v$ of the k.p model is calculated by[37]

$$c = \frac{1}{2\pi i}\left(\int_K d^2k(\partial_x A_y(k) - \partial_y A_x(k)) + \int_{K'} d^2k(\partial_x A_y(k) - \partial_y A_x(k))\right),$$

$$c_v = \frac{1}{2\pi i}\left(\int_K d^2k(\partial_x A_y(k) - \partial_y A_x(k)) - \int_{K'} d^2k(\partial_x A_y(k) - \partial_y A_x(k))\right),$$

where $A_i = \langle u_k|\partial_i|u_k\rangle$ is the berry connection. $K/K'$ represents that the interval of the integral is half of the Brillouin zone where K/K' valley locates.

# Conflict of interest

There are no conflicts to declare.

# Acknowledgements

This work is supported by the National Natural Science Foundation of China (Nos. 11804190 and 12074217), Shandong Provincial Natural Science Foundation (Nos. ZR2019QA011 and ZR2019MEM013), Shandong Provincial Key Research and Development Program (Major Scientific and Technological Innovation Project) (No. 2019JZZY010302), Shandong Provincial Key Research and Development Program (No. 2019RKE27004), Shandong Provincial Science Foundation for Excellent Young Scholars (No. ZR2020YQ04), Qilu Young Scholar Program of Shandong University, and Taishan Scholar Program of Shandong Province.